\documentclass[twocolumn,showpacs,preprintnumbers,amsmath,amssymb,floatfix,superscriptaddress,pra]{revtex4-1}
\usepackage{graphicx}
\usepackage{amsmath,amsfonts,amssymb}
\usepackage{graphicx}
\usepackage{color}
\usepackage{subcaption}

\def\beq{\begin{equation}}
\def\eeq{\end{equation}}

\newcommand{\pr}[0]{^\prime}

\newcommand{\affE}{Institute of Quantum Control (PGI-8), Forschungszentrum J\"{u}lich, D-52425 J\"{u}lich, Germany}
\newcommand{\affS}{5. Physikalisches Institut and Center for Integrated Quantum Science and Technology, Universit\"{a}t Stuttgart, Pfaffenwaldring 57, 70569 Stuttgart, Germany}

\begin{document}

\title{
Vibrational quenching of cold molecular ions immersed in their parent gas
}
\author{Krzysztof~Jachymski}\affiliation{\affE}
\author{Florian Meinert}\affiliation{\affS}

\begin{abstract}
Hybrid ion-atom systems provide an excellent platform for studies of state-resolved quantum chemistry at low temperatures, where quantum effects may be prevalent. Here we study theoretically the process of vibrational relaxation of an initially weakly bound molecular ion due to collisions with the background gas atoms. We show that this inelastic process is governed by the universal long-range part of the interaction potential, which allows for using simplified model potentials applicable to multiple atomic species.  The product distribution after the collision can be estimated by making use of the distorted wave Born approximation. We find that the inelastic collisions lead predominantly to small changes in the binding energy of the molecular ion.
\end{abstract}

\date{\today}

\maketitle

\section{Introduction.} 
%%%%%%%%%%%%%%%%%%%
%
Hybrid ion-neutral systems have been the subject of intense research~\cite{TomzaRMP}, offering the possibility to prepare highly refined molecular systems, study their dynamics in the quantum regime and possibly utilize them for quantum technological applications. As charged particles are easy to manipulate and to address with external fields, an especially promising research direction is the study of cold quantum chemistry involving ions, granting the opportunity to prepare the reactants in a well defined quantum state as well as the access to some of the reaction products. Various experimental protocols have been developed in order to bring the system to the low temperature regime, where quantum effects such as resonances can become relevant. The simplest idea relies on sympathetic cooling of the ion placed in an external trap by already ultracold atoms, which requires sufficiently large ion-to-atom mass ratio to be fully effective. Multiple experiments have been performed using this technique for an ion in a time-dependent Paul trap (see~\cite{TomzaRMP} and references therein), allowing for valuable insight in the collisional dynamics of ions in a buffer gas, including inelastic three-body recombination~\cite{Krukow2016}. A related study used the ion trap to capture the reaction products of the recombination of three neutral atoms~\cite{Wolf2017}. Recently, cooling of the system down to the $s$-wave regime where the ion-atom interaction is taking place in a single partial wave has been demonstrated~\cite{Feldker2020}. Furthermore, it has become possible to sympathetically cool ions which are placed in a static optical trap~\cite{Schmidt2019}, which in principle allows for reaching even lower temperatures. Another recently developed method is to produce an already cold ion directly from the ultracold gas with a carefully designed excitation scheme involving highly excited Rydberg states \cite{Kleinbach2018,Engel2018}. 

Reactive molecular collisions in the ultracold domain are very appealing from the theoretical point of view~\cite{Quemener2012,BohnScience17}. Initial preparation of the collision partners with very low kinetic energy and well defined internal states restricts the initially available phase space. However, formation of the collision complex opens up the possibility to rearrange the constituents and, for exoergic reactions, can lead to multiple open exit channels. In such a case, one can employ statistical methods~\cite{Bonnet1999,Forrey1999, Rackham2001,Gonzalez2014,Soley2018} to obtain some insight into the product distribution. The validity of statistical approaches relies on the assumption of the complex formation and subsequent ergodicity, which can fail for long-range processes in which the reactants do not combine into a complex, as well as for light particles for which the density of states is low. Another possibility is to disregard the quantum aspects of the process and use classical trajectory methods~\cite{Jesus2018,Jesus2019}, which proved to be accurate for ion-atom systems, as the collision partner acquire large kinetic energy while approaching each other in the presence of the attractive polarization potential. Full quantum scattering calculations are computationally very costly for multiple bodies and often have to be performed on simplified potential surfaces, but can in principle provide more complete information and test the validity of approximate methods~\cite{Bodo2006,Quemener2007,Lara2015,Stoecklin2016,Croft2017}. 

In this work, we study the collisional dynamics of a homonuclear molecular ion placed in a dilute bosonic gas of the same species. This setting is particularly relevant for experiments in which the ion is directly produced from the ultracold gas. Due to the low anisotropy of the potential surface and the lack of possible products with different nuclear arrangement, the dominant reactive process is the vibrationally inelastic collision. It turns out that the vibrational quenching is governed by long-range terms in the effective Hamiltonian, allowing for changing the internal state of the molecule without forming a three-body collision complex. Our results qualitatively agree with the classical trajectory calculations~\cite{Jesus2019}, and provide an intuitive understanding of the reaction process based on the Born approximation. We discuss the experimental parameters needed to observe the molecule dynamics and validate the theoretical predictions.

\section{Results}
%
%%%%%%%%%%%%%%%%%%%
\subsection{System}
%%%%%%%%%%%%%%%%%%%
%
Let us consider a single ion of mass $m$ moving in a dilute gas of neutral atoms of the same species. Throughout this work we will focus on the case of Rb atoms which are experimentally most relevant, but our result can also be applied to other cases. The ion motion is assumed to be unconstrained, meaning that the characteristic length scale of the ion trap (if present at all) is larger than the ion-atom interaction range as well as the typical interparticle distance. Such a situation can be realized e.g. by utilizing optical ion traps or by ionization of an ultracold atom directly from the gas. Once produced, the ion will undergo collisions with the surrounding atoms. The ion-atom interaction at long range is governed by the $-C_4/r^4$ term (leading induction coefficient in the multipole expansion)~\cite{TomzaRMP}, with $C_4=\frac{1}{2}q^2\alpha$. Here, $q$ is the ionic charge, $\alpha$ the atomic polarizability, and $r$ denotes the ion-atom distance. One can define the length scale $R^\star$ and energy scale $E^\star$ associated with this potential as
\beq
R^\star=(2\mu C_4/\hbar^2)^{(1/2)}\, , \,
E^\star=\hbar^2/2\mu (R^\star)^2\,
\eeq
with $\mu=m/2$ denoting the reduced mass of the ion-atom system. 

In addition to two-body elastic collisions and charge transfer processes which determine the ion mobility~\cite{Dalgarno1958}, three-body recombination is expected to occur, leading to association of molecular ions~\cite{Krukow2016}. The classical three-body cross section for this process has been obtained in Ref.~\cite{Jesus2018} and is described with a universal formula $\sigma_3(E)=\frac{8\pi^2}{15}2^{5/4}(R^\star)^5 (E/E^\star)^{-5/4}$. Although little is known about the product state distribution after the recombination process, experiments and theory developed for neutral Rb atoms interacting via van-der-Waals forces have led to the conclusion that the resulting molecules are mostly weakly bound~\cite{Wolf2017}. Here, rather than in the recombination, we are interested in the dynamics of the molecular ion after it has been produced. Note that it is also technically possible to associate molecular ions using a Raman process without relying on three-body recombination, which should give access to better control over the initial state~\cite{daSilvaJNP15}.
Atom-atom-ion three-body recombination has been observed in Refs.~\cite{Harter2012,Krukow2016} for trapped Rb$^+$ and Ba$^+$ ions immersed in a Rb gas prepared at typical densities of $\approx 1 \times 10^{12}~\rm{cm}^{-3}$. For the trapped ions, the typical collision energies are in the few to tens of mK regime, set by the trap-induced ion micromotion. For these conditions, recombination timescales on the order of a second have been observed. Note that for higher atomic densities typical of a Bose-Einstein condensate ($\approx 1 \times 10^{15}~\rm{cm}^{-3}$), the recombination time can be expected to reach even the microsecond timescale, highlighting the much larger cross section for ion-neutral-neutral recombination as compared to the recombination of three neutral atoms.

The homonuclear molecular ion does not posses a dipole moment, and thus its long-range interaction with the surrounding atoms is also described by the $1/r^4$ potential. The complete interaction potential surface consists of interactions among three particles, the ion (labeled as 1 in the equation below) and two atoms (labeled 2 and 3)
\beq
\label{eq:genint}
V(\mathbf{r}_{12})+V(\mathbf{r}_{13})+V(\mathbf{r}_{23})\, ,
\eeq
where $V(\mathbf{r}_{12})$ and $V(\mathbf{r}_{13})$ are the ion-atom potentials and $V(\mathbf{r}_{23})$ describes the van-der-Waals interaction between neutral atoms. For Rb, we have $R^\star=5028 \, a_0$ with $a_0$ being the Bohr radius, while the characteristic van-der-Waals length for Rb is only about $165\, a_0$. The characteristic energy $E^\star$ in the case of Rb is $h \times 1640$~Hz, or equivalently $k_B \times 78.7$~nK~\cite{TomzaRMP}.

Elastic collisions between a molecular ion and an atom do not fundamentally differ from atomic ion - neutral atom collisions, which have been widely studied already~\cite{CotePRA00,TomzaRMP}. At collision energies $E$ high enough to involve multiple partial waves, the elastic and reactive collision cross sections can be well approximated by quasiclassical formulas derived in the framework of the Langevin capture model, yielding $\sigma_{\rm el}=\pi\left(\frac{4\pi C_4^2}{\hbar^2}\right)^{1/3}\frac{1+\pi^2/16}{E^{1/3}}$ and $\sigma_{\rm re}=2\pi\sqrt{\frac{C_4}{E}}$, respectively. In the case of an initially highly excited molecular ion, a natural question to ask is how exactly will its internal state relax as a result of collisions with the buffer gas. Here the three-body nature of the problem becomes relevant. As we are only interested in processes in which the molecular ion remains bound (the collision energy is not sufficient to dissociate it), the problem can be recast as two-body scattering, but with multiple possible states after the collision, as described in the next section.

\subsection{Inelastic scattering}

Let us now briefly recall the foundations of reactive scattering theory~\cite{TaylorScattering}. First let us introduce the Jacobi coordinates and denote the relative position vector between the center of mass of the molecule and the atom by $\mathbf{R}$, and the internal coordinates of the molecular ion as $\mathbf{r}$. The center of mass motion of the particles is not relevant. At large distances where the interaction can be neglected, the molecular ion can be described by its rovibrational states $\phi_{vj}(r)$, where $v$ and $j$ label the vibrational and rotational quantum number of the molecule, respectively. It is now convenient to move to the body-fixed frame of the molecule~\cite{Curtiss1952,Arthurs1960,Jesus2016}, introducing the quantum numbers $(J,M)$ for the total angular momentum and its projection which are both conserved. The relative orbital angular momentum $\ell$ fulfills the relation $\boldsymbol{\ell}=\mathbf{J}-\mathbf{j}$. The rest of the basis states forming the basis for close-coupled equations can be labelled as $\psi_{j\ell J M}(\omega_m,\omega_{SF})$, where  $\omega_m$ and $\omega_{SF}$ are the spherical angles describing the orientation of the molecule-atom pair and the molecule itself in the space-fixed frame. 

\begin{figure}
	\centering
	\includegraphics[width=0.35\textwidth]{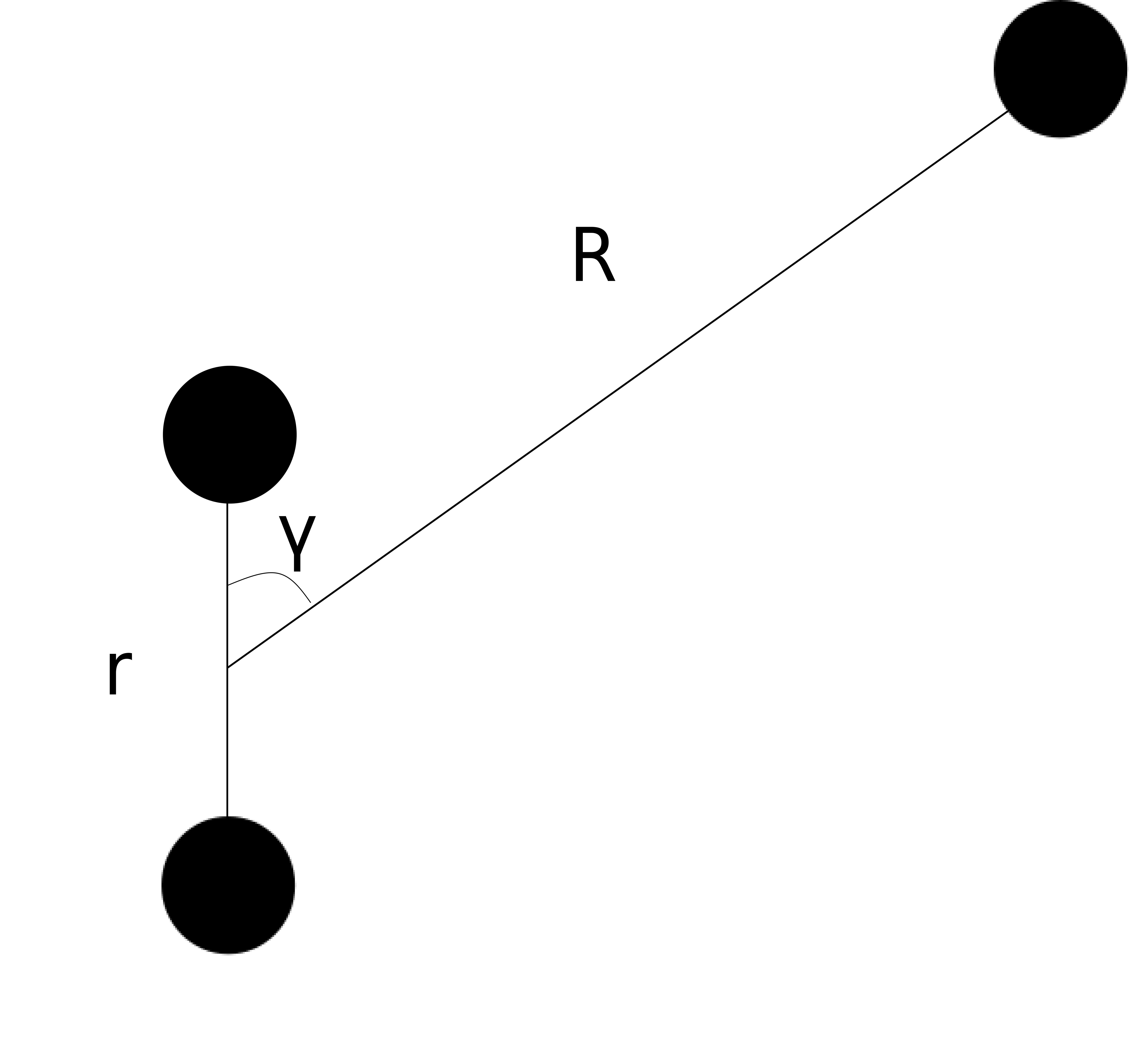}
	\caption{\label{fig:jac}Jacobi coordinates in the body-fixed frame used in this work.}
\end{figure}

We note that the three-body potential landscape is only weakly anisotropic due to the lack of a dipole moment of the molecule and the short range of the van-der-Waals interaction. This validates the use of the coupled-states approximation~\cite{McGuire1974,Pack1974,Rackham2001}, which states that the angular momentum projection $\Omega$ on the molecular axis is conserved and that in general intermultiplet couplings are negligible so that one can decouple the molecular and orbital angular momenta.

We can now calculate the potential matrix in the channel states basis simply by integrating the full three-body potential~\eqref{eq:genint} over the molecular states. The rovibrational molecular state with quantum numbers $v,\, j$ has the corresponding binding energy denoted further as $E_{vj}$. After decoupling the angular momenta, one obtains

\begin{eqnarray}
\label{eq:multich}
W^{JM} _{vj,v\pr j\pr}(R) &= &\left(\frac{2\mu^\star}{\hbar^2}E_{vj}+\frac{\ell(\ell+1)}{R^2}\right)\delta_{v v\pr}\delta_{j j\pr} +\\ \nonumber
&+& \frac{2\mu^\star}{\hbar^2}V^M _{vj,v\pr j\pr}(R)\, ,
\end{eqnarray}
with the matrix element
\begin{eqnarray}
\label{eq:matrel}
V^M  _{vj,v\pr j\pr }(R) &= &2\pi \int{ \phi_{v\pr j\pr}(r) Y_{j\pr M}(\gamma,0)} V(R,r,\gamma) \times \\ \nonumber &\times& \phi_{vj}(r) Y_{jM}(\gamma,0) \sin(\gamma) d\gamma dr
\end{eqnarray}
containing the interaction between the free atom and the molecule as the interaction term within the molecule has been incorporated in the molecular energies. Here the integration over the polar angle gives a constant factor of $2\pi$. Furthermore, $\mu^\star=2m/3$ is the reduced mass of the atom-molecular ion pair and the orbital angular momentum (which is now also a good quantum number)
\beq
\ell=\left(J(J+1)-j(j+1)-2M^2+1/4\right)^{1/2}-1/2\, .
\eeq
Furthermore, $Y_{jk}$ denote spherical harmonics and $\gamma$ is the Jacobi angle (see Figure~\ref{fig:jac}).

The nondiagonal terms of the potential matrix provide the coupling between the channels responsible for vibrationally inelastic collisions. Eq.~\eqref{eq:multich} provides the potential matrix for the multichannel scattering problem
\beq
\frac{\partial^2 \mathbf{F}}{\partial R^2}+\left(\frac{2\mu}{\hbar^2} E-\mathbf{W}(R)\right)\mathbf{F}=0
\eeq
with the matrix $\mathbf{F}$ containing the radial wavefunctions. Assuming the channels are open (as is the case if we start from the weakly bound state which has higher energy than all the others), at large distances the solution takes the canonical form
\beq
\mathbf{F}(R)\stackrel{R\to\infty}{\longrightarrow} \left(\mathbf{J}(R)-\mathbf{N}(R)\mathbf{K}\right)\mathbf{A}\, ,
\eeq
where matrices $\mathbf{J}$, $\mathbf{N}$ are diagonal and consist of spherical Bessel functions, e.g. $J_{ij}=j_\ell (k_i r)\delta_{ij}$ with $k_i=\sqrt{2\mu(E-E_i)/\hbar^2}$ being the asymptotic channel wavenumber. The amplitude matrix $\mathbf{A}$ can be linked to the reactance matrix $\mathbf{K}$ by $\mathbf{A}=\left(1-i\mathbf{K}\right)^{-1}$. Finally, the scattering matrix $\mathbf{S}$ which contains full information about the scattering is given by the formula
\beq
\label{eq:smatr}
\mathbf{S}=\left(1+i\mathbf{K}\right)\left(1-i\mathbf{K}\right)^{-1}\, .
\eeq
The inelastic collision rate constant is given by the nondiagonal elements of the S matrix and reads
\beq
\mathcal{K}_{if}=\frac{\pi \hbar}{\mu k_i}\left|S_{if}\right|^2\, .
\eeq

%
%%%%%%%%%%%%%%%%%%%
\subsection{Effective potentials}
%%%%%%%%%%%%%%%%%%%
%
With the basic formalism at hand, we now proceed to the analysis of the particular case of the molecular ion scattering with a neutral atom. The first step needed to derive the potential matrix~\eqref{eq:multich} is to calculate the two-body rovibrational wavefunctions of the molecular ion alone and their binding energies. State of the art {\it ab initio} methods can be used to calculate the interaction potential of Rb$_2^+$ in the electronic ground state with a finite precision of the order of $1\%$~\cite{Jyothi2016}. While this kind of accuracy is sufficient for spectroscopic data such as the vibrational spacing of the lower levels, it is unfortunately not enough to predict the binding energies of the weakly bound Rb$_2 ^+$ states. This is in part due to the large reduced mass of the system. For this reason we are allowed to choose a simple model potential to work with and obtain the knowledge about the general features of the ion-molecule systems. We have tested several model potentials such as the Lennard-Jones type interaction of the form $V(x)=-\frac{C_4}{x^4}(1-(x_0/x)^4)$ with the cutoff $x_0$, the flat-bottom regularization $V(x)=-\frac{C_4}{(x^2+x_0^2)^2}$ for various potential depths, as well as used the analytical solutions of the polarization potential~\cite{Idziaszek2011}, and found no fundamental difference in the behavior of the resulting effective potentials. In what follows, we will use $R^\star$ and $E^\star$ of the atom-ion system as the units of length and energy, so that asymptotically the two-body ion-atom potentials approach $-1/x^4$. 

\begin{figure*}
	\centering
	\begin{subfigure}[b]{0.32\textwidth}
		\includegraphics[width=\textwidth]{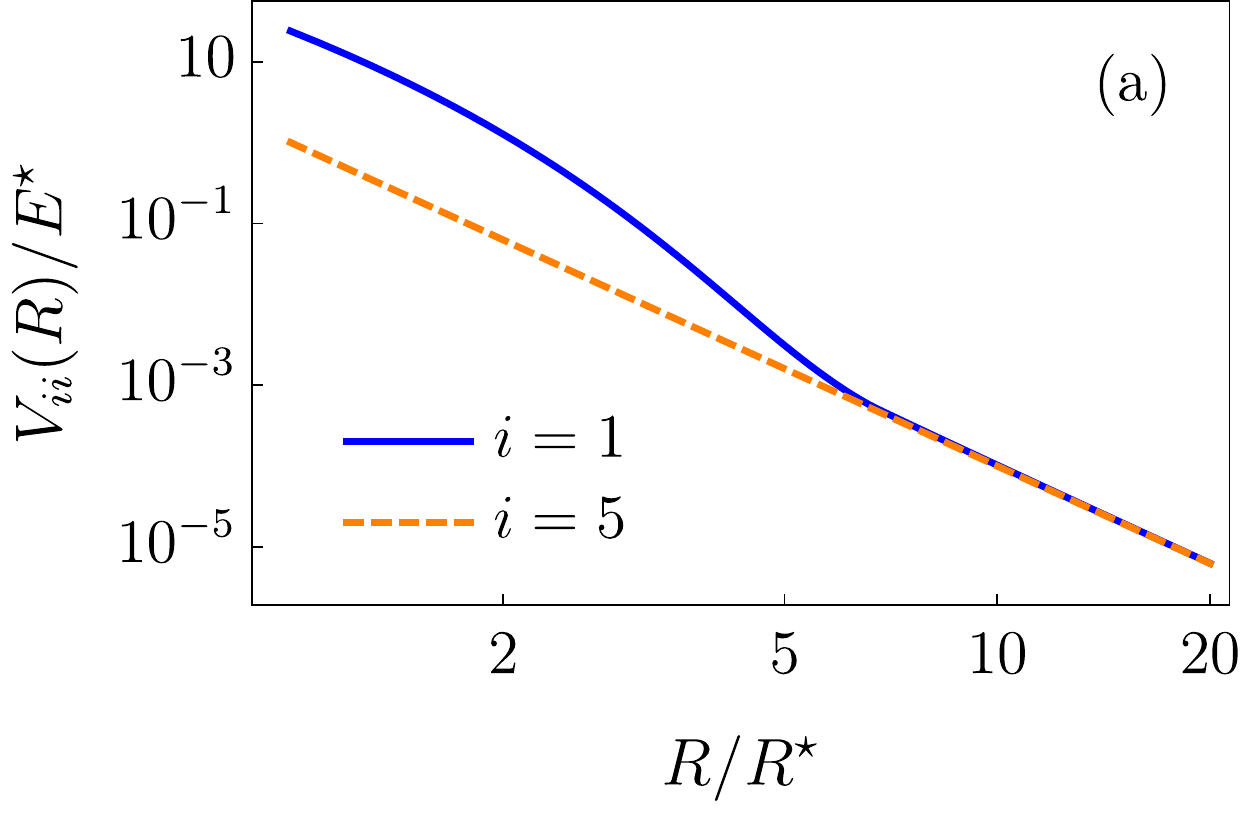}
	\end{subfigure}
	\begin{subfigure}[b]{0.32\textwidth}
		\includegraphics[width=\textwidth]{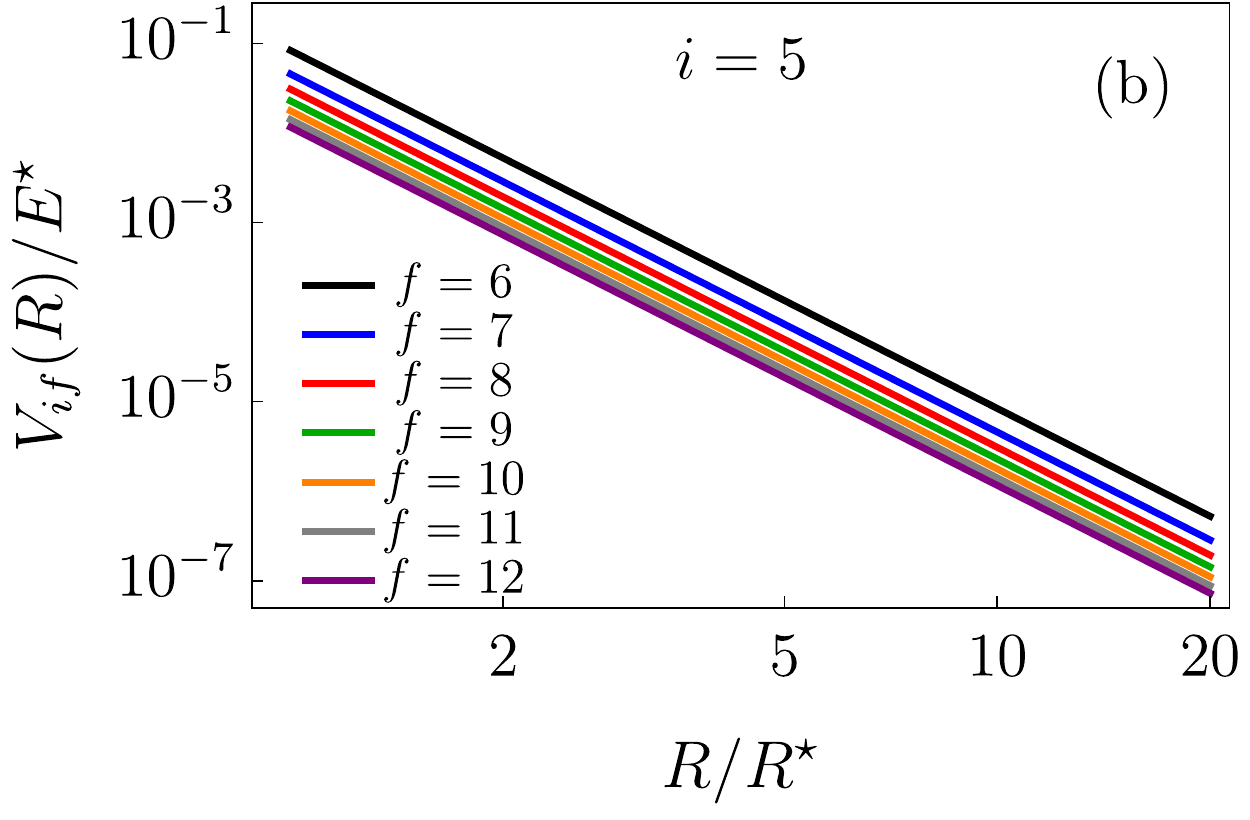}
	\end{subfigure} 
	\begin{subfigure}[b]{0.32\textwidth}
		\includegraphics[width=\textwidth]{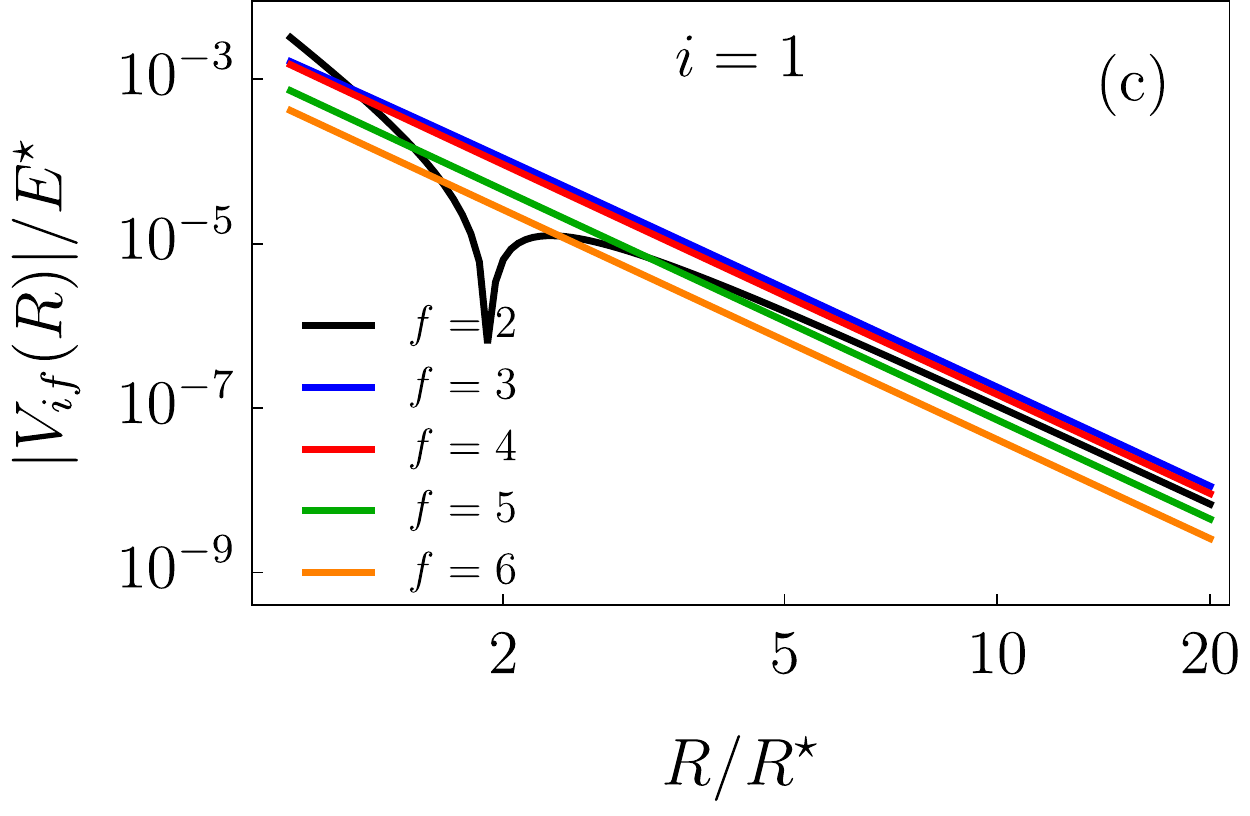}
	\end{subfigure}
	\caption{(a) Diagonal effective potentials for the most weakly bound state (blue) and the 5th one counting from the dissociation threshold (orange).  (b) Couplings between the 5th vibrational state and the ones right below it as a function of the atom-molecule distance. (c) Same couplings calculated for the most weakly bound state. The binding energies $E_v\approx 2\cdot10^4\,E^\star$ for $v=5$ and $0.16\,E^\star$ for $v=1$.}
	\label{Fig_Vii}
\end{figure*}

One might expect that the effective atom-molecule potentials given by Eq.~\eqref{eq:matrel} will follow the $R^{-4}$ power law, but with a possibly varying coupling coefficient. This can be understood assuming the deeply bound states are pointlike compared to the range of the interparticle potential, and the contribution to the integral~\eqref{eq:matrel} comes only from $r\sim 0$.
The next term which can be extracted perturbatively and takes into account the spatial structure of the molecule can be obtained by expanding the angle-averaged atom-molecule potential $V_{\rm at-mol}=V(\mathbf{r}_{13})+V(\mathbf{r}_{23})$ at large distances $R$ in power series, which gives 
\beq
V_{\rm at-mol}\stackrel{R\to\infty}{\longrightarrow}-\frac{1}{R^4}+\frac{r^2/2+R_6^6}{R^6}+\dots\, ,
\eeq
with $R_6$ denoting the characteristic distance corresponding to the van-der-Waals interaction $R_6=(2\mu C_6/\hbar)^{1/4}$, which is typically an order of magnitude smaller than $R^\star$~\cite{JulienneRMP10}. Thus, there is a contribution to the coupling matrix element proportional to $\int{dr\,\phi_{vj}(r)\phi_{v\pr j\pr}(r) \frac{r^2}{R^6}}$. This term should be relevant for weakly bound states whose spatial extent cannot be neglected, both for the diagonal and nondiagonal matrix elements. This intuitively explains why the potential curves are largely independent of the exact form of the two-body potentials, as the short-range corrections only set in at distances smaller than $R^\star$.

Two of the diagonal matrix elements of Eq.~\eqref{eq:matrel} are presented in the first panel of Fig.~\ref{Fig_Vii}. As expected, they decay according to the $R^{-4}$ power law at long range with the fitted $C_4$ coefficient equal to the original one. For the most weakly bound state, which in our case extends to around $10\,R^\star$ and has the binding energy of only $\approx0.16\,E^\star$, the van-der-Waals term is also clearly visible. However, as this only provides a short-range correction we can conclude that the elastic scattering is just governed by the polarization potential with some unknown scattering length which depends on the short-range details.

\begin{figure}
	\centering
	\begin{subfigure}[b]{0.4\textwidth}
		\includegraphics[width=\textwidth]{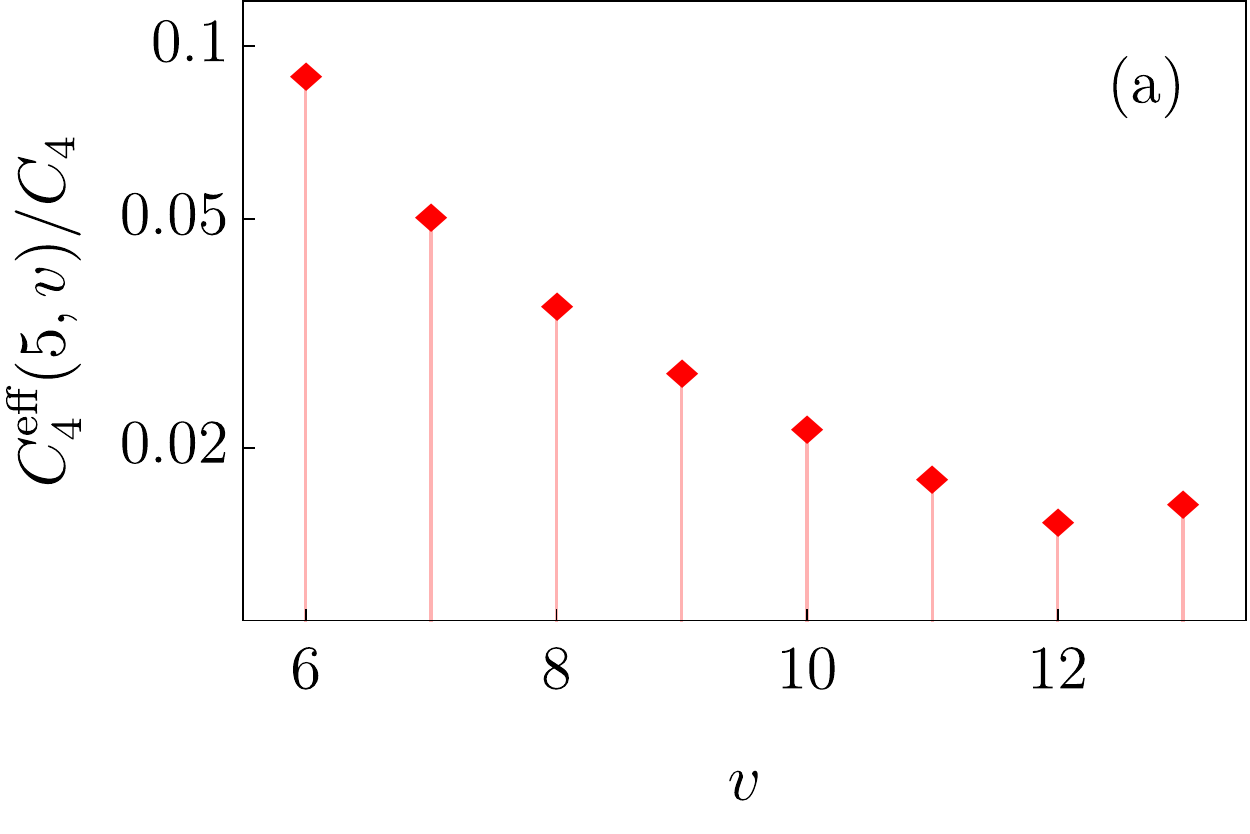}
	\end{subfigure}
	\begin{subfigure}[b]{0.4\textwidth}
		\includegraphics[width=\textwidth]{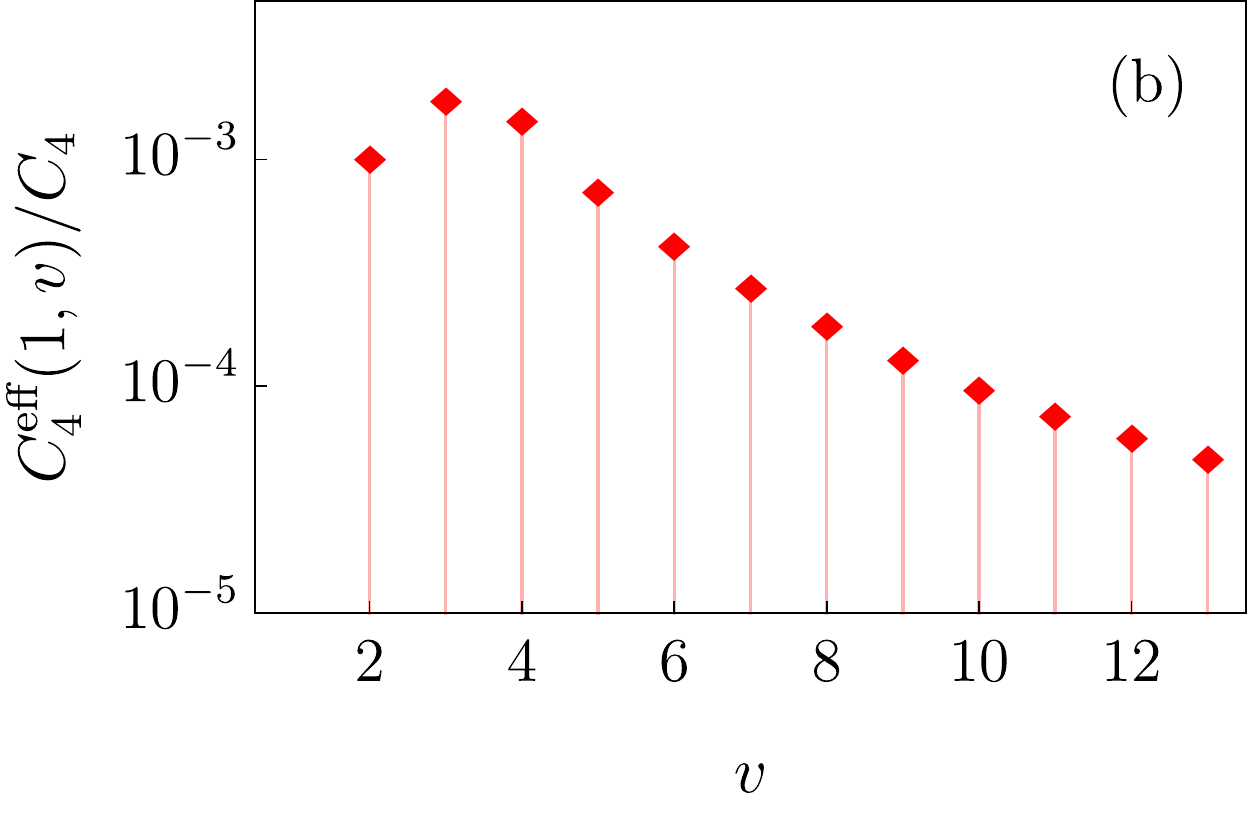}
	\end{subfigure}
	\caption{(a) Effective coupling coefficients $C_4^{\rm eff}$ between the 5th vibrational state and the ones right below it in units of the diagonal coupling constant. (b) Same, but for the most weakly bound state.}
	\label{Fig_C4eff}
\end{figure}

Two exemplary cases of the coupling potential curves are shown in the middle and the right panel of Fig.~\ref{Fig_Vii}. As expected, the couplings between deeply bound levels (middle panel) always follow the effective power-law potential with $1/R^4$ dependence. However, for the most weakly bound state (right panel) the coupling to the state right below it has a different form as we again observe the impact of the $1/R^6$ term at $R\lesssim 5R^\star$. In addition, the potential changes its sign, leading to a dip on the logarithmic scale that has a finite depth due to the data points resolution. The effective $C_4$ coupling coefficients generally decrease for states lying further from each other, as summarized in Fig.~\ref{Fig_C4eff}.

The calculations shown here have been performed for the ground rotational state of the molecular ion. We have checked that the excited rotational states show the same asymptotic behavior leading to $\sim 1/R^4$ dependence of the couplings. As we are dealing with the case of homonuclear ions, the rotationally inelastic collisions are suppressed. This is due to the small anisotropy of the potential surface in the absence of the charge-dipole term, which would be present for a heteronuclear molecular ion giving a contribution $V_{\rm c-d}\propto\cos(\gamma)/R^5$~\cite{Jesus2016}. In our case the leading terms in the potential that couple different rotational levels behave like $r^2/R^6$, being relevant only in higher order.

Finally, let us mention that for vibrational relaxation the impact of the atom-atom interaction in the potential landscape is negligible due to the length scale separation between the van-der-Waals and polarization potential. The van-der-Waals forces set in only at distances at which the particles would already form a collision complex, which is beyond the scope of the current analysis. 

%
%%%%%%%%%%%%%%%%%%%
\subsection{Distorted wave Born approximation}
%%%%%%%%%%%%%%%%%%%
%
Having calculated the effective interaction potential, we now turn to the analysis of the scattering. At this point, an issue concerning the short-range effects arises. Namely, if the reactants form a collision complex, the short-range three-body physics becomes relevant and impacts the recombination process. Ergodic dynamics of the complex would lead to statistical distribution of the product states. The effective potential~\eqref{eq:multich} is written in the rovibrational state basis and thus does not capture dynamics of the three-body complex. Furthermore, as demonstrated above, the coupling between different vibrational states of the molecular ion arising in the collision has rather long-range nature. It is then warranted to focus on the close-coupled system described by Eq.~\eqref{eq:multich}, disregarding the possibility of the complex formation. A possible minimal extension of the current results would be to impose (partially) absorbing boundary conditions at short range in order to mimick the long-lived complex creation and treat the part of the flux reaching the inner region statistically~\cite{Rackham2001,Jachymski2014}.

Rather than solving the scattering problem set by Eq.~\eqref{eq:multich} in full, as we are anyway working with an approximate model potential, we will now proceed by employing the distorted wave Born approximation (DWBA) to estimate the inelastic collision rates. DWBA is a method based on perturbation theory~\cite{TaylorScattering,ChildMolecular}, which utilizes the knowledge of the scattering solution to the elastic part of the problem and has been widely used to describe atom-molecule inelastic collisions~\cite{Miller1968,Forrey1999,Volpi2002}. This is particularly suitable for our problem since the diagonal potentials are analytically solvable~\cite{Idziaszek2011}.

Within the DWBA, the off-diagonal elements of the $\mathbf{K}$ matrix in Eq.~\eqref{eq:smatr} are given by
\beq
\label{eq:kmatr}
K_{if}=-\pi\int{dR\, u_i(R)u_f(R)V_{if}(R)}\, ,
\eeq
where the $u_i$ functions denote the energy normalized scattering solution for the diagonal potential in channel $i$ (in our case given solely by the $1/R^4$ term) and $V_{if}$ is the coupling matrix element between the initial and the final state given by Eq.~\eqref{eq:matrel}. The elastic scattering within the DWBA is not affected by the nondiagonal terms in the potential and can be described analytically by energy- and angular momentum-dependent phase shifts $K_{ii}=\tan\delta(\ell_i,k_i)$. Then the scattering matrix and all observable quantities can be calculated from Eq.~\eqref{eq:smatr}. Note that the matrix elements need to be calculated for all partial waves involved in the collision.

Interestingly, assuming that the coupling between the channels behaves as $1/R^4$, the matrix element~\eqref{eq:kmatr} can be calculated analytically assuming the channel wavefunctions are given by spherical Bessel functions and for angular momenta of the two states fulfilling $\ell+\ell\pr\geq 1$. This provides the threshold law for the K-matrix element $K\propto k_i^{\ell_i+1/2}$. In particular, for $\ell_i=0$ one obtains $K_{if}=2k_i^{1/2}k_f^{3/2}\phantom{I}_2 F_1((-1-\ell_f)/2,(-1+\ell_f)/2,1,k_i^2/k_f^2)/(\ell_f^2-1)$ where $\phantom{I}_2 F_1$ is the hypergeometric function. In the limit of a far-from-threshold exit channel (large energy difference between the states compared to the initial kinetic energy) this quantity scales as $k_f^{3/2}$. In practice, we are anyway imposing a short-range cutoff to the interaction and use the full scattering solutions of the polarization potential which oscillate quickly at small distances giving no contribution there. 

\begin{figure}
	\centering
	\includegraphics[width=0.4\textwidth]{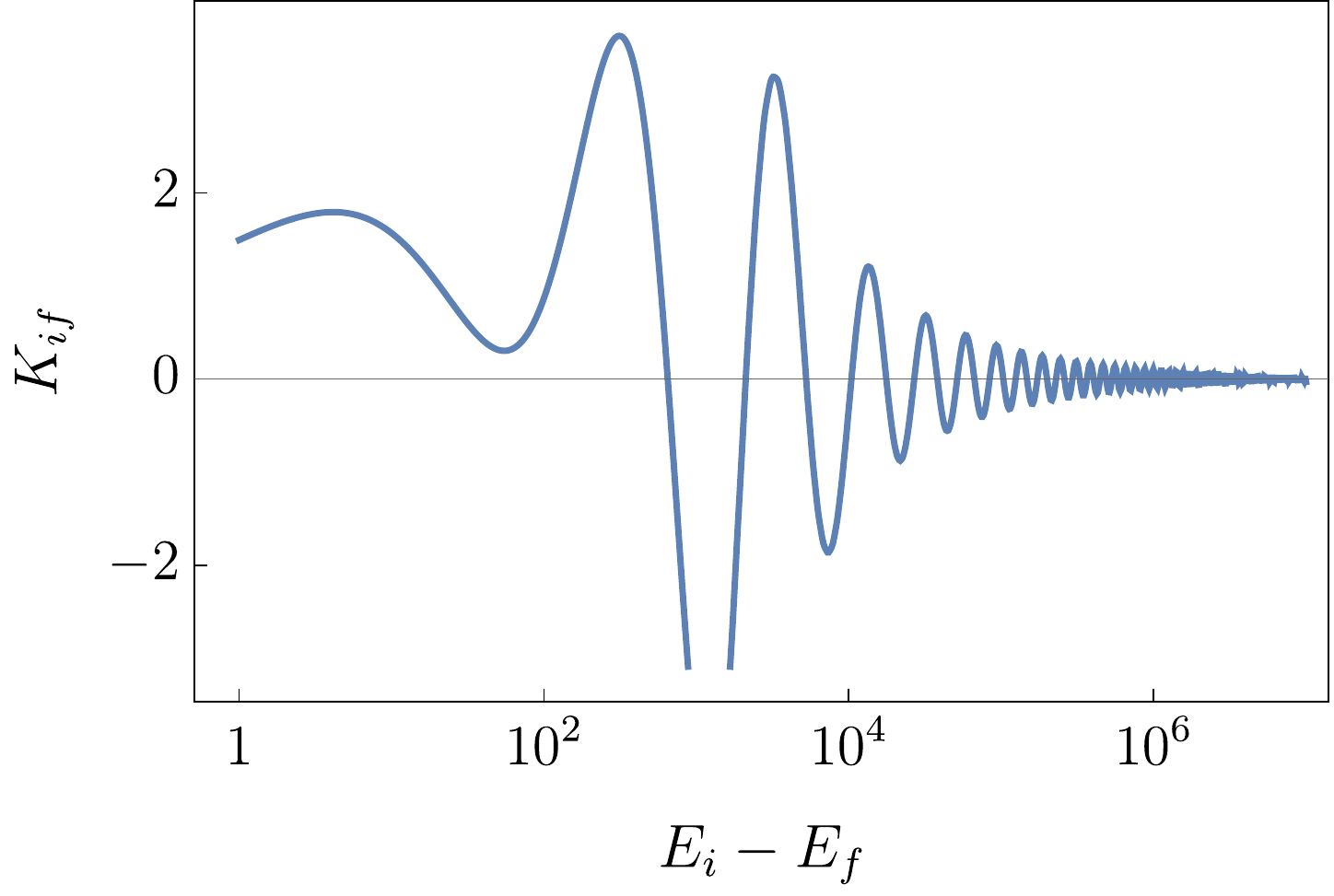}
	\caption{K matrix elements calculated within the DWBA for vanishing collision energy as a function of the energy difference between the channels for the entrance channel scattering length $a=R^\star$ and initial kinetic energy $E=10^{-4}E^\star$.}
	\label{FigKmat}
\end{figure}

In Fig.~\ref{FigKmat}, we show the dependence of the inelastic K-matrix element (assuming a dimensionless $1/R^4$ potential) on the energy difference between the two channels for the initial state with kinetic energy of $10^{-4}\,E^\star$ and scattering length equal to $R^\star$. The rapid decrease in magnitude and fast oscillations result from the properties of the outgoing state. At relatively low energy difference between the states (roughly below $10^4\,E^\star$) one could in principle expect visible resonant effects provided that the reactants can be prepared in a single quantum state (although when the calculated matrix elements are large, DWBA cannot be trusted any more). However, the polarization potential anyway does not support a dense enough bound-state spectrum and apart from the coupling between the two most weakly bound levels one cannot expect any resonant enhancement. Moreover, in experiment the results would be further blurred by thermal averaging. More specifically, in order to calculate the rate constant at finite temperature one needs to take a convolution according to $\mathcal{K}(T)= \frac{2}{\sqrt{\pi}(k_B T)^{3/2}}\int{dE\,\mathcal{K}(E)E^{1/2}e^{-E/k_B T}}$.

\begin{figure}
	\centering
	\begin{subfigure}[b]{0.4\textwidth}
		\includegraphics[width=\textwidth]{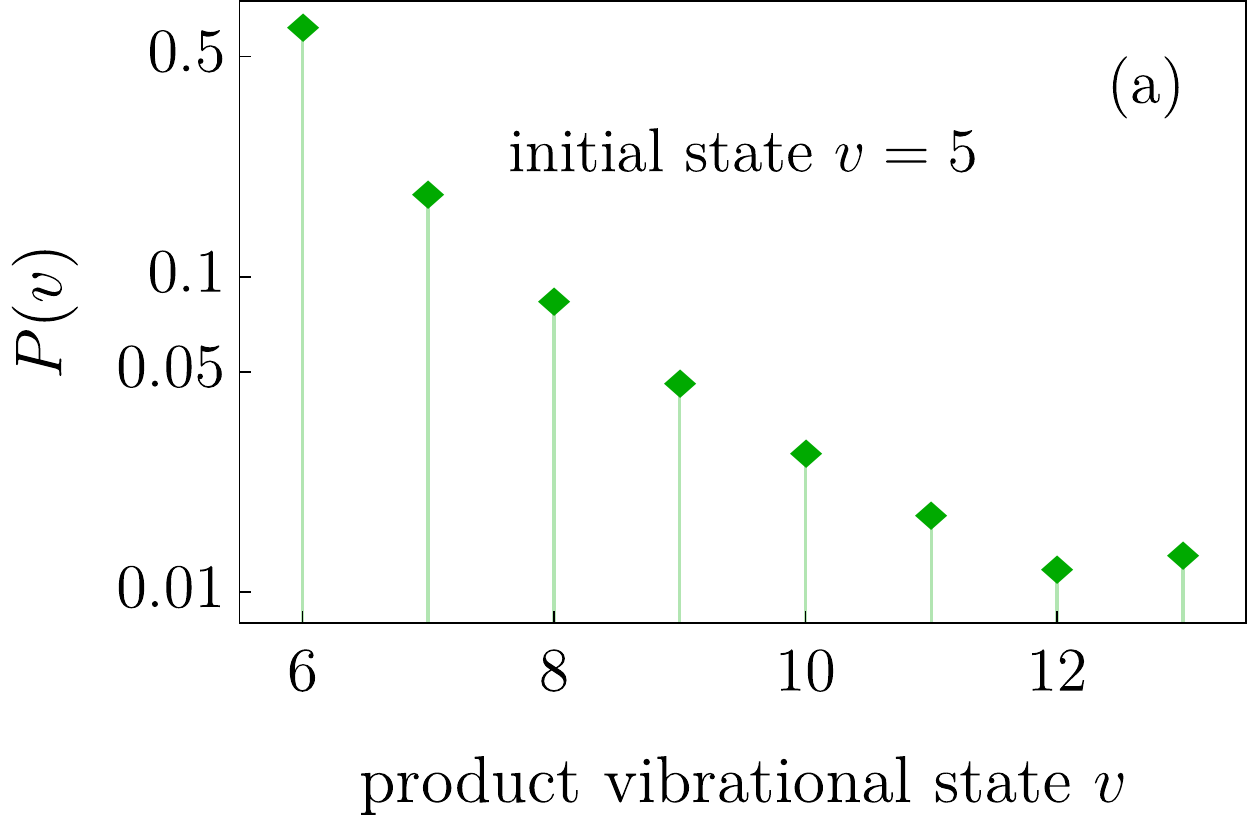}
	\end{subfigure}
	\begin{subfigure}[b]{0.4\textwidth}
		\includegraphics[width=\textwidth]{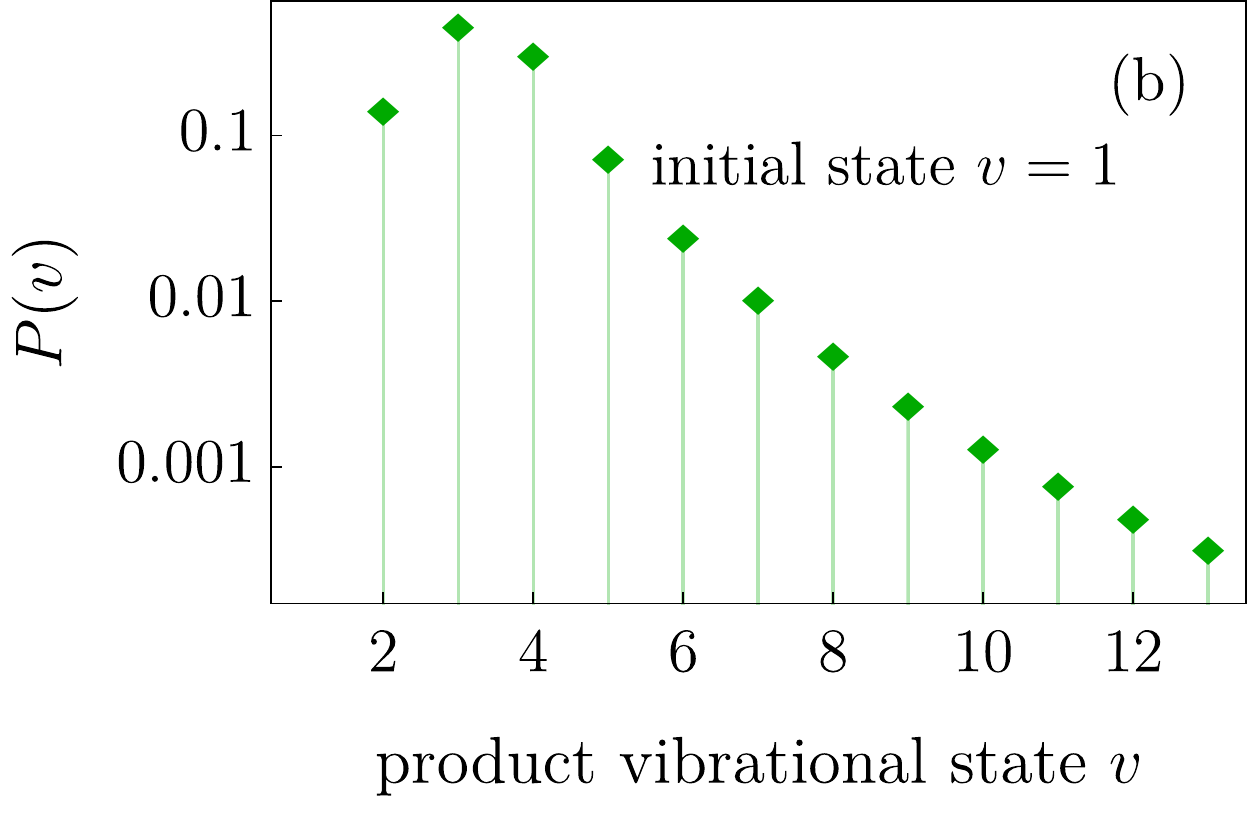}
	\end{subfigure}
	\caption{(a) Distribution of product states for vanishing collision energy after the inelastic collision for the 5th vibrational state ($v=5$). (b) Same, but for the most weakly bound state ($v=1$).}
	\label{FigProduct}
\end{figure}

Along with the coupling coefficients calculated above, the DWBA allows for a simple estimate of the product distribution. As an example, we focus again on the initial vibrational state $v=5$ and $v=1$ counting from the dissociation threshold. The obtained distribution is shown in Fig.~\ref{FigProduct}. The result of the collision is affected by the magnitude of the effective coupling coefficients as well as the difference in energy between the channels. We have averaged the results over the rapid oscillations of the Bessel functions involved in the integral~\eqref{eq:kmatr}. We find that in both cases the distribution of reaction products is rather broad and favors the nearby states, in agreement with the classical trajectory calculations~\cite{Jesus2019}. Due to the large number of available product states, the total inelastic collision cross section follows the universal Langevin prediction $\sigma_{\rm re}=2\pi\sqrt{\frac{C_4}{E}}$. We have checked that as long as the collision energy $E\lesssim 10^3 E^\star$ ($\approx k_B \times 100\mu$K for Rb) the product distribution does not change significantly.

%
%%%%%%%%%%%%%%%%%%%
\section{Discussion}
%%%%%%%%%%%%%%%%%%%
%
We have analyzed the process of vibrational quenching of an initially excited homonuclear molecular ion due to collisions with the background gas. The effective couplings between the molecular states turn out to decay with a power law in the same way as the polarization potential, but with different coupling coefficients depending on the overlap of the molecular wave functions. The long-range nature of the process makes it possible to rely on distorted wave Born approximation to determine the distribution of the collision products. This approximate quantum treatment agrees qualitatively with previous estimates based on classical trajectory calculations.

The total inelastic rate coefficient for vibrational relaxation of the weakly bound molecular ions is given by the Langevin formula $\mathcal{K}_L=2\pi\hbar R^\star_{\rm{mol}}/\mu^\star$ with $R^\star_{\rm{mol}}$ being the characteristic length of the molecular ion-atom pair, irrespective of the initial vibrational state. Assuming the case of Rb atoms, $\mathcal{K}_L \approx 2.1\cdot10^{-9}$cm$^3/$s, which for high atomic densities $n\approx 10^{15}$cm$^{-3}$, typical for a Bose-Einstein condensate, results in lifetimes of the order of microseconds. In order to estimate the role of such secondary relaxation processes after an initial three-body recombination event for experiments, we consider a typical kinetic energy release of $\approx h \times (10 \, ... \, 100)$~MHz associated with the formation of a weakly bound molecular ion. Evidently, this yields a mean free path of the molecular ion on the order of few micrometers between successive collision events, significantly smaller than the typical spatial extent of a trapped condensate. This means that in a dense gas secondary collisions after the production of the molecular ion can be highly important and can be detected e.g. by monitoring the arrival time of the molecule at the detector. For comparison, in the case of neutral weakly bound Rb$_2$ molecule the expected rate coefficient is 50 times smaller.

In our model we neglected the role of possible complex formation. Once all three particles involved come close together, the resulting complex state may be long-lived and provide high density of resonance states. This would presumably impact the result of the collision by redistributing the particles along all possible product states in a statistical way. Extension of our model to involve three-body resonances and complex formation will be the subject of future work.

{\it Acknowledgements}. F.M. acknowledges support from Deutsche Forschungsgemeinschaft [Project No. PF 381/17-1, part of the SPP 1929 (GiRyd)], funding by the Carl-Zeiss foundation, and is indebted to the Baden-W\"{u}rttemberg-Stiftung for the financial support by the Eliteprogramm for Postdocs.

\bibliography{giant_ion}

\end{document}